\documentclass[11pt]{article}
\usepackage{moriond,epsfig}

\def\Journal#1#2#3#4{{#1} {\bf #2}, #3 (#4)}

\def\NIMA{{\em Nucl. Instrum. Methods} A}
\def\NPB{{\em Nucl. Phys.} B}
\def\PLB{{\em Phys. Lett.}  B}
\def\PRL{\em Phys. Rev. Lett.}
\def\PRD{{\em Phys. Rev.} D}

\def\mco{\multicolumn}

\def\be{\begin{equation}}
\def\ee{\end{equation}}
\def\bea{\begin{eqnarray}}
\def\eea{\end{eqnarray}}

\begin{document}
\vspace*{4cm}
\title{DI-BOSON PRODUCTION AT THE TEVATRON}

\author{ GILLES DE LENTDECKER \\
On behalf of the CDF and D$\emptyset$ collaborations}

\address{Department of Physics and Astronomy,\\
 University of Rochester, Rochester, 14627 NY, USA}

\maketitle\abstracts{
We present some precision measurements on electroweak physics performed at the Tevatron collider at Fermilab. Namely we report on the boson-pair production cross sections and on triple gauge boson couplings using proton anti-proton collisions collected by the CDF and D$\emptyset$ experiments at the center-of-mass energy of 1.96~TeV. The data correspond to an integrated luminosity of up to 324~pb$^{-1}$.
}

\section{Introduction}

The CDF~\cite{CDF} and D$\emptyset$~\cite{D0} experiments 
study proton anti-proton collisions at $\sqrt{s} = 1.96$~TeV at the Tevatron collider at Fermilab (Batavia, IL, USA). The Run~II of the Tevatron (started in 2001) gives us a unique opportunity to study $W$ and $Z$ boson physics. Compared to the LEP and SLD accelerators, where many of the best measurements of electroweak parameters have been made, the Tevatron offers the advantage to produce a larger number of $W$ bosons and to produce $Z/\gamma^*$ at higher invariant masses.

\par
Study of events containing pairs of vector bosons produce important tests of the non-Abelian structure of the Standard Model (SM). The $SU(2)_L \times U(1)_Y$ structure of the SM implies that the electroweak gauge bosons $W$ and $Z$ can interact with one another through trilinear and quartic gauge boson vertices. Non-SM values of these couplings may increase the di-boson production cross section significantly. Therefore a measurement of this quantity provides a sensitive test of the SM and probes for low energy remnants of new physics.

In this note we review some new results on $W^+ W^-$, $WZ$, $ZZ$, $W\gamma$ and $Z\gamma$ production cross section as well as limits on $WWZ$, $WW\gamma$, $ZZ\gamma$ and $Z\gamma\gamma$ anomalous couplings.

\section{$W$ and $Z$ signatures}
Due to a large QCD background, decay channels involving quarks are difficult to measure; therefore $W$ and $Z$ bosons are mainly identified through their leptonic decays. These decays are characterized by a high transverse energy ($E_T$) lepton and large transverse missing 
energy~($E_T$\hspace*{-0.45cm}{/}\hspace*{0.45cm}) for $W$, or by two high 
transverse energy leptons for $Z$. Typically the lepton $E_T$ is required to be
greater than 20~-~25~GeV and $E_T$\hspace*{-0.45cm}{/}\hspace*{0.45cm}  
greater than 20~-~25~GeV as well.

\section{$W^+ W^-$ production cross section}
The first evidence of $W$ pair production was found in $p \bar{p}$ collisions by the CDF Collaboration at $\sqrt{s} = 1.8$~TeV~\cite{WWRI}. The properties of $W$ pair production have been extensively studied by the LEP collaborations~\cite{LEP} but the Tevatron offers the possibility to probe much higher masses.
\begin{table}[htb]
\caption{Number of expected signal ($N_{sig}$) and background ($N_{bg}$) events and number of observed events ($N_{obs}$) for the three dilepton channels. The integrated luminosity is also given. Systematic uncertainties are included for CDF, not for D$\emptyset$.}
\label{table:ww}
\vspace{0.4cm}
\begin{center}
\begin{tabular}{|c|c|c|c|c|c|c|}
\hline
& \mco{3}{|c|}{D$\emptyset$} & \mco{3}{|c|}{CDF} \\ \hline 
 & $ee$ & $\mu \mu$ & $e \mu $ & $ee$ & $\mu \mu$ & $e \mu$ \\ \hline
Lum (pb$^{-1}$)& 252 & 224 & 235 & 184 & 184 & 184 \\
$N_{sig}$ & 3.42~$\pm$~0.05 & 2.10~$\pm$~0.05 & 11.10~$\pm$~0.1 & 2.6~$\pm$~0.3 & 2.5~$\pm$~0.3 & 5.1~$\pm$~0.6\\
$N_{bg}$  & 2.30~$\pm$~0.21 & 1.95~$\pm$~0.41 & 3.81~$\pm$~0.17 & 1.9~$^{+1.3}_{-0.3}$ & 1.3~$^{+1.6}_{-0.4}$ & 1.9~$\pm$~0.4 \\
$N_{obs}$ &  6 & 4 & 15 & 6 & 6 & 5 \\
\hline
\end{tabular}
\end{center}
\end{table}

\begin{figure}
\setlength{\unitlength}{1.0cm}
\begin{center}
\vskip 2.5cm
\begin{minipage}[t]{7.0cm}
\begin{picture}(3.5,3.5)
\put(-1.0,-1.0){\psfig{figure=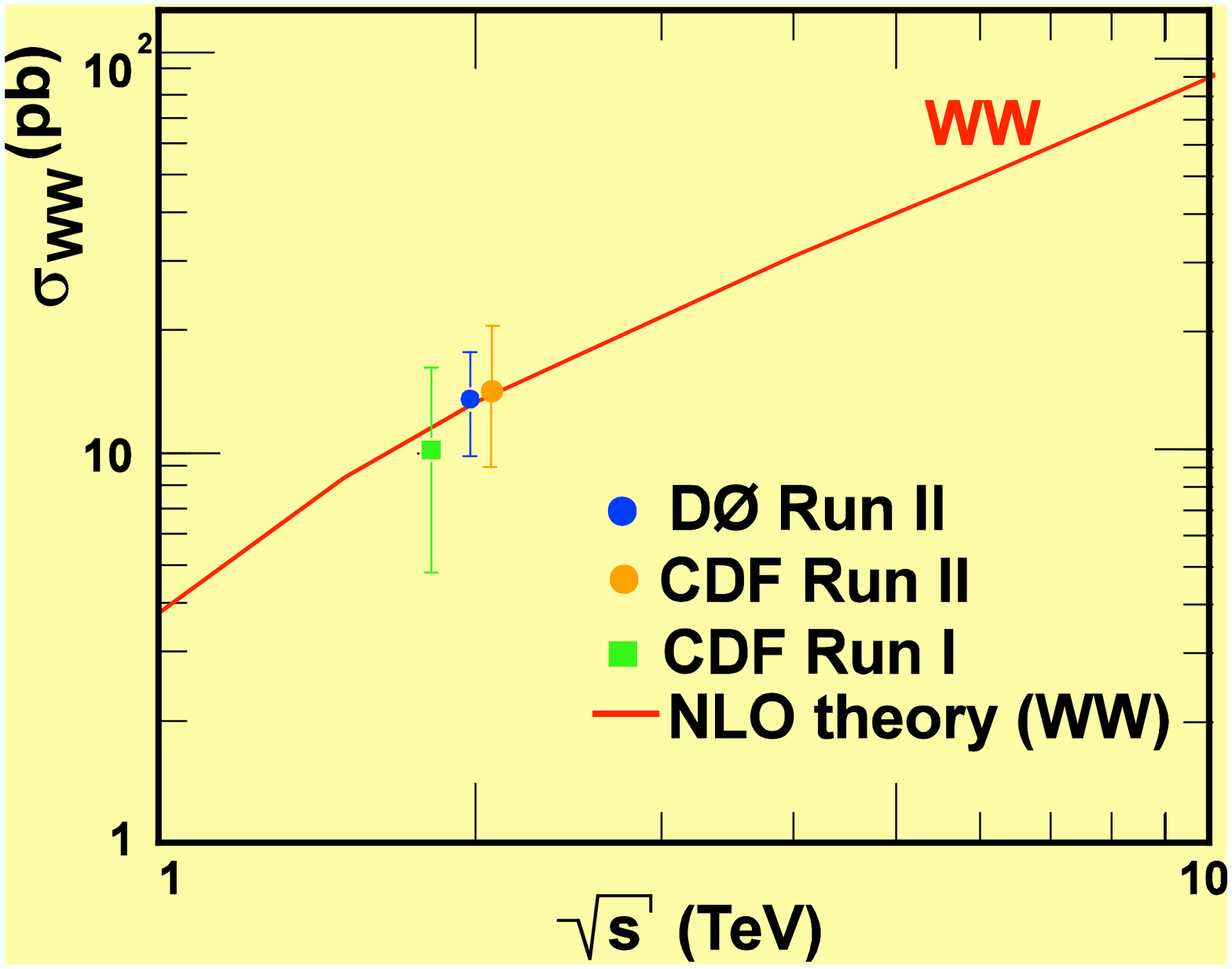,height=7.0cm}}
\end{picture}
\caption{$\sigma (p \bar{p} \rightarrow W^+ W^-)$ as a function of $\sqrt{s}$.
The experimental measurements are compared with the NLO calculation.
\label{fig:ww}}
\end{minipage}
\hfill
\begin{minipage}[t]{8.0cm}
\begin{picture}(3.5,3.5)
\put(-0.5,0.0){\epsfig{figure=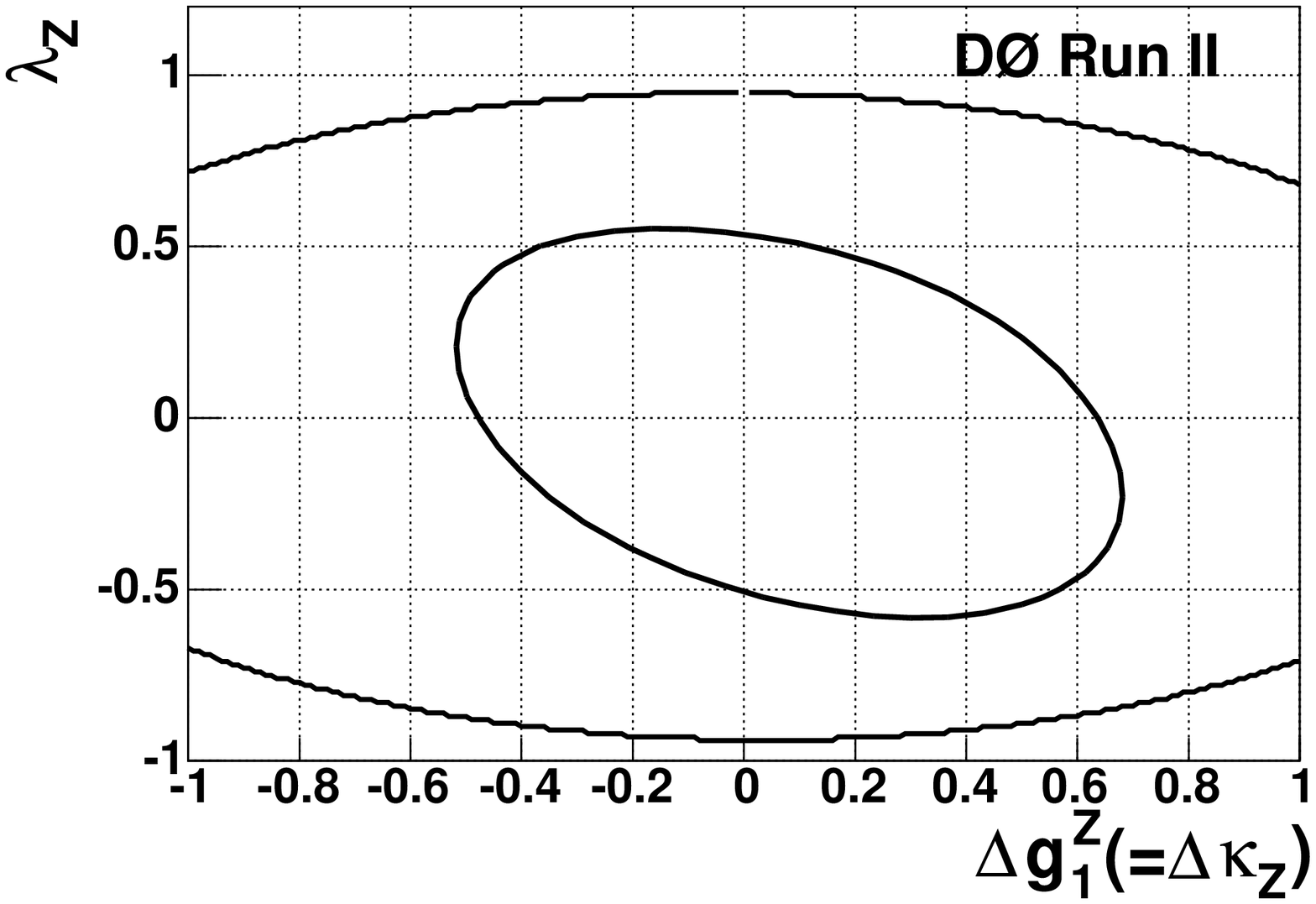,height=6.0cm}}
\end{picture}
\caption{Two-dimensional coupling limits at 95\% C.L (inner contour) on 
$\lambda_Z$ vs $\Delta g^1_Z$ for $\Lambda~=~1.5$~TeV obtained by D$\emptyset$
with 285-320~pb$^{-1}$ of data. The outer contour 
corresponds to the unitarity limit. 
\label{figure:wwz_2d_limits}}
\end{minipage}
\end{center}
\end{figure}

In this note we present the measurement of $W^+ W^-$ production cross section in the dilepton decay channel $W^+ W^- \rightarrow l^+ \nu l^- \nu$  ($l = e, \mu$). Candidate events are required to have two well identified, oppositely charged, leptons (electrons or muons). Significant backgrounds to $W^+ W^-$ production in the dilepton channels include Drell-Yan events with large $E_T$ \hspace*{-0.6cm}{/}\hspace*{0.25cm}, 
$t \bar{t}$ production and $WZ$, $ZZ$ production. The number of expected signal
($N_{sig}$) and background ($N_{bg}$) events and the number of observed events 
($N_{obs}$) are given in Table~\ref{table:ww} for CDF and D$\emptyset$, together
with the integrated luminosity for each decay channel~\cite{ww_cdf}~\cite{ww_d0}. As a final result, 
the combined cross section for $WW$ production at a center-of-mass energy of $\sqrt{s} = 1.96$~TeV is :
\begin{equation}
\begin{array}{rcl}
\sigma (p \bar{p} \rightarrow W^+ W^-) & = & 14.6^{+5.8}_{-5.1}~(\rm{stat})~^{+1.8}_{-3.0}~(\rm{syst})~\pm~0.9~(lum)~\rm{pb~(CDF)} \nonumber \\
& = & 13.8^{+4.3}_{-3.8}~(\rm{stat})~^{+1.2}_{-0.9}~(\rm{syst})~\pm~0.9~(lum) ~\rm{pb~(D\emptyset)} \nonumber \\
\end{array}
\end{equation}
These values are in good agreement with the NLO calculation of 12.4~$\pm$~0.8~pb~\cite{MCFM}, as shown in Fig.~\ref{fig:ww}. D$\emptyset$ also calculates the 
probability that the observed events are caused by a fluctuation of the 
background. This probability amounts to $2.3 \times 10^7$, corresponding to 5.2 standard deviations.

\section{$WZ$ and $ZZ$ pair production}
The presence of unexpected neutral triple gauge boson couplings ($ZZZ$ and 
$ZZ\gamma$) can result in an enhanced rate of $ZZ$ production, and a anomalous
$WWZ$ coupling can increase the $WZ$ production rate above the SM prediction.
In general the $WWV$ ($V=Z,\gamma$) interactions can be described by an 
effective Lagrangian with arbitrary parameters $g^1_V$, $\lambda_V$ and 
$\kappa_V$~\cite{HAG87}. In 
the SM $g^1_Z = \kappa_Z = 1$ and $\lambda_Z = 0$. To avoid unitarity 
violation at high energies, it is necessary to introduce form factors with 
scale $\Lambda$.

A test for anomalous trilinear boson couplings using $WZ$ events is unique in 
that $WZ$ diagrams contain only $WWZ$ and not $WW\gamma$ vertices. Anomalous 
trilinear gauge boson coupling limits produced using $W^+ W^-$ events are 
sensitive to both vertices and must make an assumption about the relation of 
the $WWZ$ to the $WW\gamma$ coupling. See for instance the HISZ 
relations~\cite{HISZ93}. As this analysis is performed using $W^{\pm}Z$ event
candidates that are unavailable at $e^+e^-$ colliders, it provides a unique 
measurement of $WWZ$ anomalous coupling limits.

The cleanest $WZ$ signals consist of final states with three charged leptons and a neutrino. Requiring three isolated high transverse momentum ($p_T$) leptons 
and large $E_T$\hspace*{-0.45cm}{/}\hspace*{0.25cm} associated with the 
neutrino strongly 
suppresses all known SM backgrounds. The main background comes from $Z+X$ 
events, where $X$ is a jet, a photon or a $Z$.

\begin{table}[htb]
\caption{Number of signal ($N_{sig}$) and background ($N_{bg}$) events expected and number of events observed ($N_{obs}$) by the D$\emptyset$ experiment for 
the four tri-lepton channels. The integrated luminosity is also given.}
\label{table:wz}
\vspace{0.4cm}
\begin{center}
\begin{tabular}{|c|c|c|c|c|}
\hline
Decay Channels &  $N_{obs}$ & $N_{sig}$ & $N_{bg}$ & lum (pb$^{-1}$)\\ \hline
$eee$          & 1 & 0.44~$\pm$~0.07 & 0.155~$\pm$~0.043 & 320\\ 
$ee \mu$       & 0 & 0.45~$\pm$~0.04 & 0.073~$\pm$~0.029 & 292\\
$\mu \mu e$    & 0 & 0.53~$\pm$~0.06 & 0.349~$\pm$~0.034 & 285\\
$\mu \mu \mu$  & 2 & 0.62~$\pm$~0.08 & 0.132~$\pm$0.053 & 289\\ \hline
Total          & 3 & 2.04~$\pm$~0.13 & 0.71~$\pm$~0.08   & $-$\\ \hline
\end{tabular}
\end{center}
\end{table}

The number of observed candidates ($N_{obs}$), the number of expected signal 
($N_{sig}$) and background ($N_{bg}$) events are given in Table~\ref{table:wz} for the D$\emptyset$ Collaboration for each trilepton channel~\cite{wz_d0}. 
The 2.04~$\pm$~0.13 expected $WZ$ events combined with the 0.71~$\pm$~0.08 
estimated background events are consistent with the three candidate events 
found by D$\emptyset$. The probability for a background to fluctuate to three 
or more candidates is 3.5\%. The corresponding
$\sigma (p \bar{p} \rightarrow WZ)$ amounts to $4.5^{+3.8}_{-2.6}$~pb and the 95\% confidence level (C.L.) upper limit is $< 13.3$~pb.

\begin{table}[htb]
\caption{The expected contributions from SM $ZZ$, $ZW$ and background sources in 194~$pb^{-1}$ of CDF data.}
\label{table:wz_zz}
\vspace{0.4cm}
\begin{center}
\begin{tabular}{|c|c|c|c|c|}
\hline
Process & 4 leptons & 3 leptons & 2 leptons & combined \\ \hline
$ZZ$    & 0.06~$\pm$~0.01 & 0.13~$\pm$~0.01 & 0.69~$\pm$~0.11 & 0.88~$\pm$~0.13\\
$WZ$    &        -        & 0.78~$\pm$~0.06 & 0.65~$\pm$~0.10 & 1.43~$\pm$~0.16\\ \hline
$N_{sig}$ &  0.06~$\pm$~0.01 & 0.91~$\pm$~0.07 & 1.34~$\pm$~0.21 & 2.31~$\pm$~0.29 \\ \hline
$WW$ & $ - $ & $ - $ & 0.40~$\pm$~0.07 & 0.40~$\pm$~0.07 \\
Fake & 0.01~$\pm$~0.02 & 0.07~$\pm$~0.06 & 0.21~$\pm$~0.12 &  0.29~$\pm$~0.16 \\
Drell-Yan & $ - $ & $ - $ & 0.31~$\pm$~0.17 & 0.31~$\pm$~0.17 \\
$t \bar{t}$  & $ - $ & $ - $ & 0.02~$\pm$~0.01 & 0.02~$\pm$~0.01 \\ \hline
$N_{bg}$ & 0.01~$\pm$~0.02 & 0.07~$\pm$~0.06 & 0.94~$\pm$~0.22 & 1.02~$\pm$~0.24 \\ \hline
$N_{sig} + N_{bg}$ & 0.07~$\pm$~0.02 & 0.98~$\pm$~0.09 & 2.28~$\pm$~0.35 & 3.33~$\pm$~0.42 \\ \hline
$N_{obs}$ & 0 & 0 & 3 & 3 \\ \hline
\end{tabular}
\end{center}
\end{table}

CDF performed a combined analysis to measure the $WZ$ and $ZZ$ cross sections~\cite{wz_cdf}. 
They studied events in three categories designed to encompass the main leptonic
ratios of the $WZ$ and $ZZ$ decays. The first include events with four charged 
leptons, which is sensitive to $ZZ \rightarrow lll'l'$ ($l=e, \mu, \tau$). The second category includes events with three charged leptons plus 
$E_T$\hspace*{-0.45cm}{/}\hspace*{0.45cm}. It consists predominantly of 
$ZW \rightarrow lll' \nu$. Events from $ZZ \rightarrow lll'l'$, where one 
lepton is not identified, can also fall in this category. The third category 
includes events with two charged leptons plus 
$E_T$\hspace*{-0.45cm}{/}\hspace*{0.45cm}. This category is sensitive to 
$ZZ \rightarrow ll\nu \nu$ and $ZW \rightarrow lll' \nu$, where one lepton is 
not identified. The number of observed candidates ($N_{obs}$) and the number 
of expected signal ($N_{sig}$) and background ($N_{bg}$) events are given in 
Table~\ref{table:wz_zz} for each category. The main background in the 
four-lepton and three-lepton categories is from 'fake-lepton' events, in which 
jets have been mis-identified as leptons in $Z/W$+jets events. The backgrounds
in the two-lepton category include $WW$, $t \bar{t}$, Drell-Yan and 
fake-lepton events. The expected numbers of the 
different backgrounds are shown in Table~\ref{table:wz_zz}. 
Having insufficient statistics to measure the cross section, CDF sets a 95\% 
C.L. upper limit on $\sigma 
(p \bar{p} \rightarrow WZ~+~ZZ)$ of  $< 15.2$~pb, which is in agreement with 
the NLO SM expectation of $\sigma (p \bar{p} \rightarrow WZ~+~ZZ)~=~5.0~\pm~0.4$~pb. The probability for the background of 1.02~$\pm$~0.24 to fluctuate to 
give three or more events is 9\%.

\begin{table}[htb]
\caption{One-dimensional 95\% C.L. intervals on $\lambda_Z$, $g^1_Z$ and $\kappa_Z$ obtained with 285-320~pb$^{-1}$ of D$\emptyset$ data.}
\label{table:wwz_limits}
\vspace{0.4cm}
\begin{center}
\begin{tabular}{|c|c|}
\hline
$\Lambda~=~1$~TeV & $\Lambda~=~1.5$~TeV   \\ \hline
$-0.53~<~\lambda_Z~<~0.56$ & $-0.48~<~\lambda_Z~<~0.48$ \\
$-0.57~<~\Delta g^1_Z~<~0.76$ & $-0.49~<~\Delta g^1_Z~<~0.66$ \\
$-2.0~<~\Delta \kappa_Z~<~2.4$ & $ - $ \\ \hline
\end{tabular}
\end{center}
\end{table}

Using the three observed trilepton candidates, D$\emptyset$ sets one- and two-dimensional limits on $\lambda_Z$, $\Delta g^1_Z$ and $\Delta \kappa_Z$, where
$\Delta g^1_Z \equiv g^1_Z - 1$ and $\Delta \kappa_Z \equiv \kappa_Z - 1 $.
Table~\ref{table:wwz_limits} lists one-dimensional 95\% C.L. limits on these coupling constants for two form factor scale $\Lambda~=~1$~and 1.5 TeV. The experimental limit on 
$\Delta \kappa_Z$ exceeds the unitarity limit with $\Lambda~=~1.5$~TeV. 
Fig.~\ref{figure:wwz_2d_limits} shows 
the two-dimensional 95\% C.L. contour limits with $\Lambda~=~1.5$~TeV. 
These results are the tightest limits on anomalous $WWZ$ couplings derived from $WZ$ final state.


\section{$W\gamma$ production}\label{section:wg}
The $W\gamma$ production can be used to study the $WW\gamma$ vertex. The final state which is used in this study is $p \bar{p} \rightarrow l \nu \gamma$ 
($l = e, \mu$). 
In the SM this final state occurs due to $W\gamma \rightarrow l \nu \gamma$ as well as via lepton bremsstrahlung $W \rightarrow l \nu \rightarrow l \nu \gamma$. The $W$ selection is the same as in the $W$ inclusive cross section measurement and is described elsewhere in these proceedings~\cite{DELIOT05}. In addition to the $W$, a high-energy ($E^{\gamma}_T~>~7$~GeV) photon isolated from the lepton ($\Delta R(\gamma-l)~>~0.7$) is required to suppress events with final state radiation of the photon from the outgoing lepton.
The dominant background is $W$~+~jet production where a jet mimics a photon.

\begin{table}[htb]
\caption{Number of expected signal ($N_{sig}$) and background ($N_{bg}$) events
and number of observed events ($N_{obs}$) for $e \nu \gamma$ and  $\mu \nu \gamma$ production for D$\emptyset$ and CDF. The first cross section uncertainty is 
statistical and the second is systematic.}
\label{table:wgamma}
\vspace{0.4cm}
\begin{center}
\begin{tabular}{|c|c|c|c|c|}
\hline
& \mco{2}{|c|}{D$\emptyset$} & \mco{2}{|c|}{CDF} \\ \hline 
 & $e \nu \gamma$ &  $\mu \nu \gamma$ & $e \nu \gamma$ &  $\mu \nu \gamma$ \\ \hline
Lum (pb$^{-1}$) & 162 (6.5\%) & 134 (6.5\%) & 168 - 202 (6\%) & 175 - 192 (6\%) \\
$N_{sig}$ & 51.2~$\pm$~11.5 & 89.7~$\pm$~13.7 & 126.8~$\pm$~5.8 & 95.2~$\pm$~4.9 \\
$N_{bg}$ & 60.8~$\pm$~4.5 & 71.3~$\pm$~5.2 & 67.3~$\pm$~18.1 & 47.3~$\pm$~7.6 \\
$N_{obs}$ & 112 & 161 & 195 & 128 \\ \hline
$\sigma(l \nu \gamma) (\rm{pb})$ & 13.9~$\pm$~2.9~$\pm$~1.6 & 15.2~$\pm$~2.0~$\pm$~1.1 & 19.4~$\pm$~2.1~$\pm$~2.9 & 16.3~$\pm$~2.3~$\pm$~1.8 \\ \hline 
\end{tabular}
\end{center}
\end{table}

The number of observed candidates ($N_{obs}$) and the number of expected signal ($N_{sig}$)  and background ($N_{bg}$) events are given in Table~\ref{table:wgamma} for the CDF and D$\emptyset$ collaborations for the electron and muon channels~\cite{wg_cdf}~\cite{wg_d0}. The resulting cross sections are also given in Table~\ref{table:wgamma}. Combining both channels, assuming lepton universality and taking into account correlations of the systematic uncertainties, yields $\sigma (l \nu \gamma)~=~18.1~\pm~3.1$~pb for CDF. The theoretical prediction for this cross section and for the kinematic region
$E^{\gamma}_T~>~7$~GeV and $\Delta R(\gamma-l)~>~0.7$ is 19.3~$\pm$~1.4~pb.
The combined cross section of D$\emptyset$ is measured to be $\sigma (l \nu \gamma)~=~14.8~\pm~1.6~(\rm{stat})~\pm~1.0~(\rm{syst})~\pm~1.0~(\rm{lum})$~pb, in agreement with the SM prediction of 16.0~$\pm$~0.4~pb, for the kinematic region 
$E^{\gamma}_T~>~8$~GeV and $\Delta R(\gamma-l)~>~0.7$.

\begin{figure}
\setlength{\unitlength}{1.0cm}
\vskip 2.5cm
\begin{center}
\begin{picture}(3.5,4.5)
\put(-6.5,0.0){\epsfig{figure=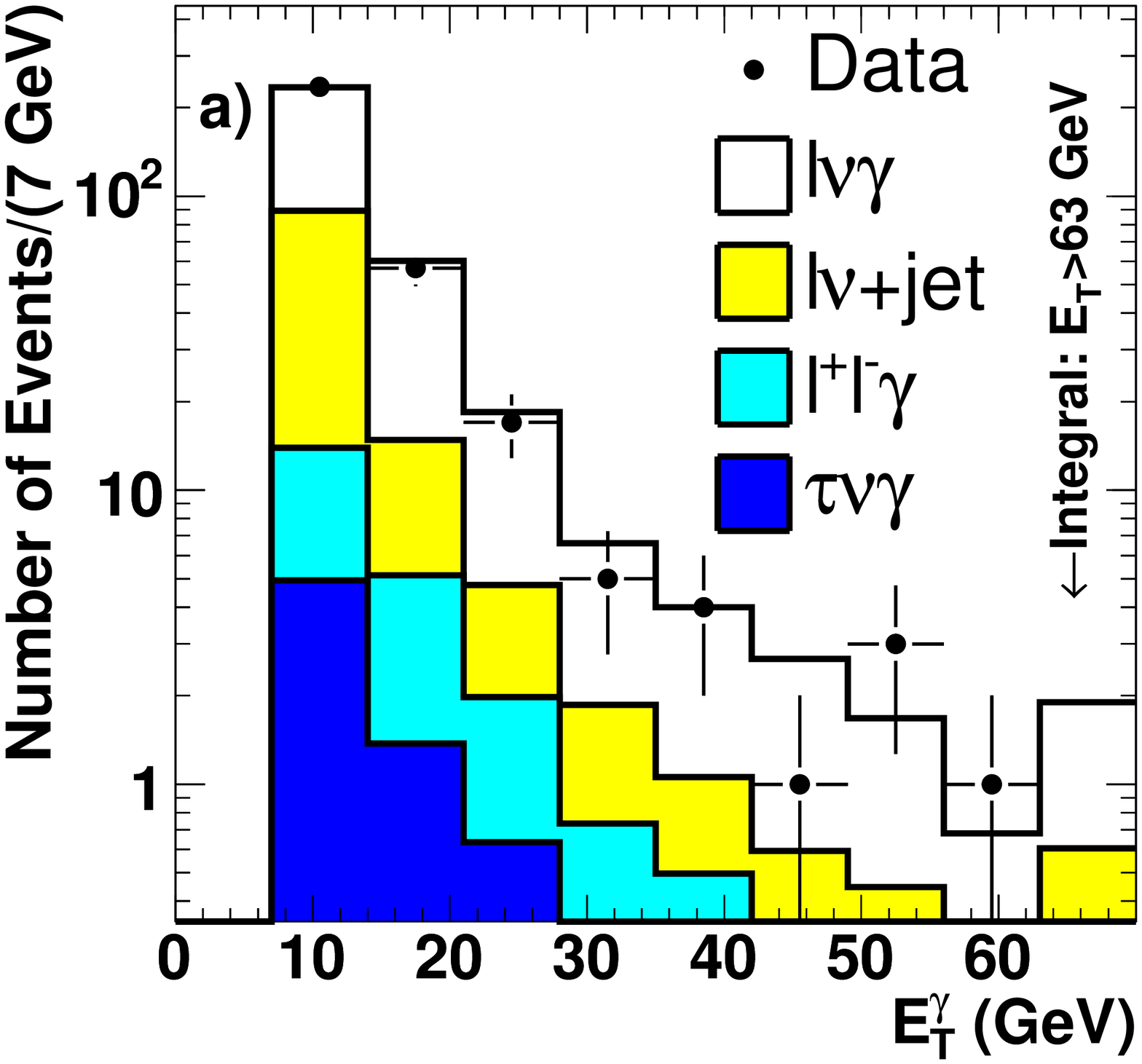,height=6.0cm}}

\put(1.0,0.0){\epsfig{figure=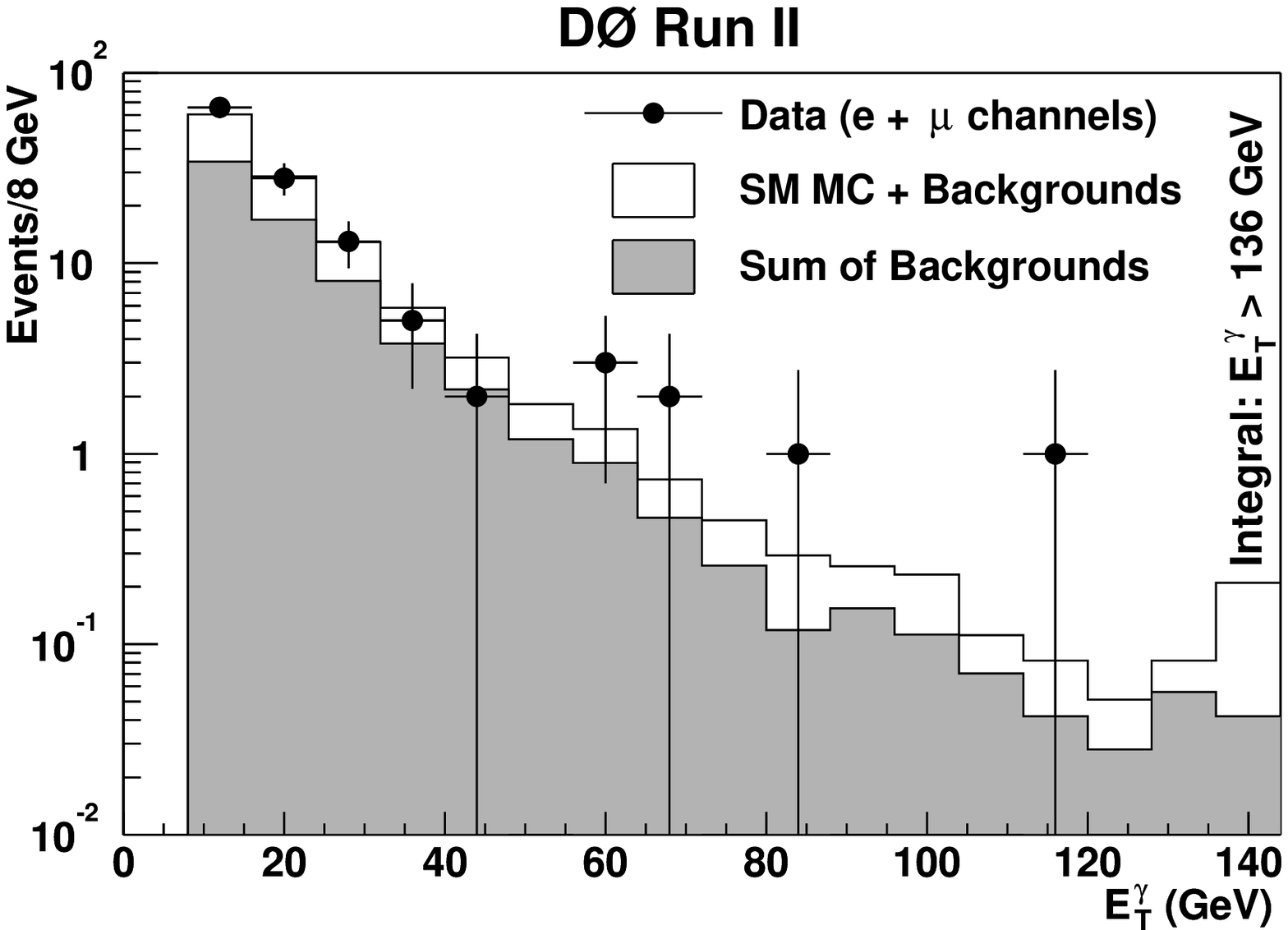,height=6.0cm}}
\end{picture}
\end{center}
\caption{The photon $E_T$ spectrum of the candidate events for the $W \gamma$ candidates for CDF (left) and D$\emptyset$ (right). The data (black points) are 
compared with the SM expectations for signal and backgrounds (open histogram).
\label{figure:wg_spectra}}
\end{figure}

\begin{figure}
\vskip 1.5cm
\begin{center}
\epsfig{figure=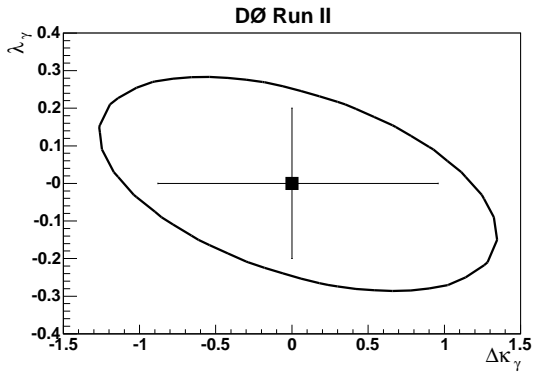,height=5.0cm}
\end{center}
\caption{One- and two-dimensional 95\% C.L. limits on $WW\gamma$ coupling parameters $\Delta \kappa_{\gamma}$ and $\lambda_{\gamma}$ using 134-162~pb$^{-1}$ of
D$\emptyset$ data.
\label{figure:wwg_limits}}
\end{figure}

Fig.~\ref{figure:wg_spectra} shows the photon $E_T$ spectrum for CDF (left) and D$\emptyset$ (right). The data (black points) are compared with the SM expectations for signal and backgrounds (open histogram). The background estimates are indicated with shaded histograms. The data are in good agreement with the SM expectations, and no enhancement of the photon $E_T$ spectrum is seen at high 
transverse energy. D$\emptyset$ sets one- and two-dimensional 95\% C.L. limits
on the coupling parameters $\Delta \kappa_{\gamma}$ and $\lambda_{\gamma}$
(see Fig.~\ref{figure:wwg_limits}).
The one-dimensional limits on each parameter are 
$-0.93 < \Delta \kappa_{\gamma} < 0.97$ and 
$-0.22 < \lambda_{\gamma} < 0.22$. These limits 
represent the most stringent constraints on anomalous $WW\gamma$ couplings 
obtained by direct observation of $W\gamma$ production.

\section{$Z \gamma$ production.}
In the SM the trilinear gauge couplings of the $Z$ boson to the photon are 
zero. Therefore photons do not interact with Z bosons at lowest order. 
Evidence of such an interaction would indicate new physics. 
The study of $Z$ boson and photon production is a stepping stone for the 
analysis of $ZZ^*\gamma$ and $Z\gamma^*\gamma$ trilinear gauge couplings. 

We present the study of $Z \gamma$ production using $Z$ boson decays to $e^+e^-$ and $\mu^+ \mu^-$. The photon may be emitted through initial state radiation (ISR) from one of the partons or produced as final state radiation (FSR) from one of the final leptons. The SM $Z \gamma$ production processes produce photons with a rapidly falling transverse energy $E^{\gamma}_T$. In contrast anomalous 
$ZZ^*\gamma$ or $Z\gamma^*\gamma$ couplings which appear in extensions of the 
SM, can cause production of photons with high $E^{\gamma}_T$ and can increase 
the $l^+l^-\gamma$ cross section compared to the SM prediction. 
In the formalism described in Ref.~\cite{BAUR93} one assumes that the 
$ZV\gamma$ ($V=Z,\gamma$) couplings are Lorentz and gauge-invariant. The most 
general $ZV\gamma$ coupling is parameterized by two CP-violating 
($h^V_{10}$ and $h^V_{20}$) and two CP-conserving ($h^V_{30}$ and $h^V_{40}$) complex 
coupling parameters. The $Z\gamma$ candidate events are selected using the same
selection as in the $Z$ inclusive cross section measurement (described elsewhere in these proceedings~\cite{DELIOT05}) but with the identification of a photon as described in section~\ref{section:wg}. The main background to $Z\gamma$ is $Z$+jet production, where the jet is misidentified as a photon.

\begin{table}[htb]
\caption{Number of expected signal ($N_{sig}$) and background ($N_{bg}$) events and number of observed events ($N_{obs}$) for $e e \gamma$ and  $\mu \mu \gamma$ production for {D$\emptyset$} and CDF.}
\label{table:zgamma}
\vspace{0.4cm}
\begin{center}
\begin{tabular}{|c|c|c|c|c|}
\hline
& \mco{2}{|c|}{D$\emptyset$} & \mco{2}{|c|}{CDF} \\ \hline 
 & $e e \gamma$ &  $\mu \mu \gamma$ & $e e \gamma$ &  $\mu \mu \gamma$ \\ \hline
Lum (pb$^{-1}$) & 324 (6.5\%) & 286 (6.5\%) & 168 - 202 (6\%) & 175 - 192 (6\%) \\
$N_{sig}$ & 109~$\pm$~7 & 128~$\pm$~8 & 31.3~$\pm$~1.6 & 33.6~$\pm$~1.5 \\
$N_{bg}$ & 23.6~$\pm$~2.3 & 22.4~$\pm$~3.0 & 2.8~$\pm$~0.9 & 2.1~$\pm$~0.7 \\
$N_{obs}$ & 138 & 152 & 36 & 35 \\ \hline 
\end{tabular}
\end{center}
\end{table}

The number of observed candidates ($N_{obs}$) and the number of expected signal ($N_{sig}$)  and background (N$_{bg}$) events are given in Table~\ref{table:zgamma} for the CDF and D$\emptyset$ collaborations for the electron and muon channels~\cite{wg_cdf}~\cite{zg_d0}. The resulting cross sections of CDF are 
$\sigma(e^+e^-\gamma)$~=~4.8~$\pm$~0.8~(stat)~$\pm$~0.3~(syst)~pb and 
$\sigma(\mu^+\mu^-\gamma)$~=~4.4~$\pm$~0.8~(stat)~$\pm$~0.2~(syst)~pb.
The combined cross section measured by CDF is 
$\sigma(l^+l^-\gamma)$~=~4.6~$\pm$~0.6~pb in agreement with the theoretical prediction including the photon acceptance of $4.5~\pm~0.3$~pb~\cite{BAUR96}. 
The combined cross section measured by D$\emptyset$ is 
$\sigma(l^+l^-\gamma)$~=~4.2~$\pm$~0.4~(stat + syst)~$\pm$~0.3~(lum)~pb in agreement with the theoretical prediction including the photon acceptance of $3.9^{+0.1}_{-0.2}$~pb~\cite{BAUR98}.

\begin{figure}
\setlength{\unitlength}{1.0cm}
\begin{center}
\vskip 2.5cm
\begin{picture}(15.0,4.0)
\put(1.0,0.0){\epsfig{figure=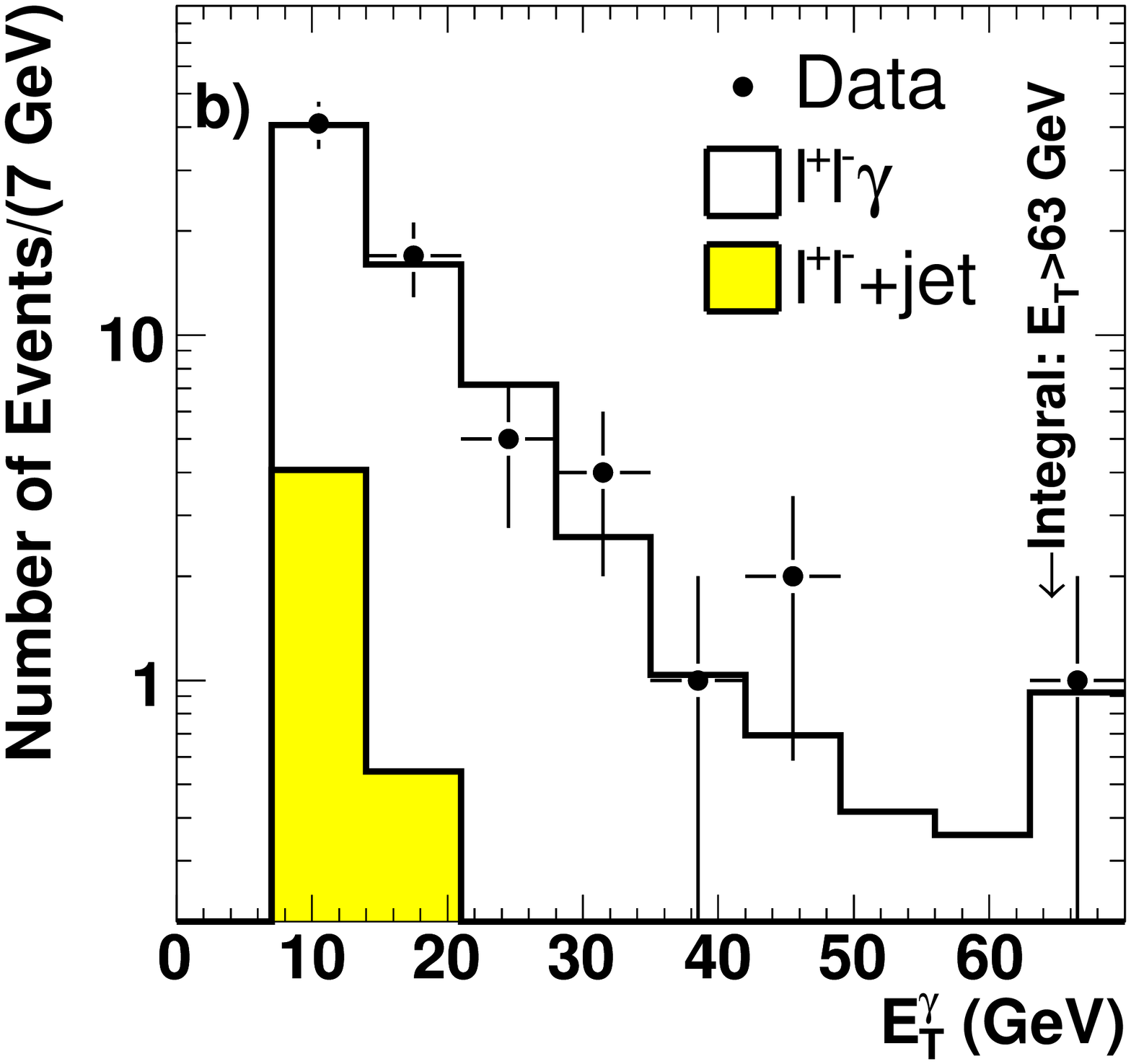,height=6.0cm}}

\put(8.0,0.0){\epsfig{figure=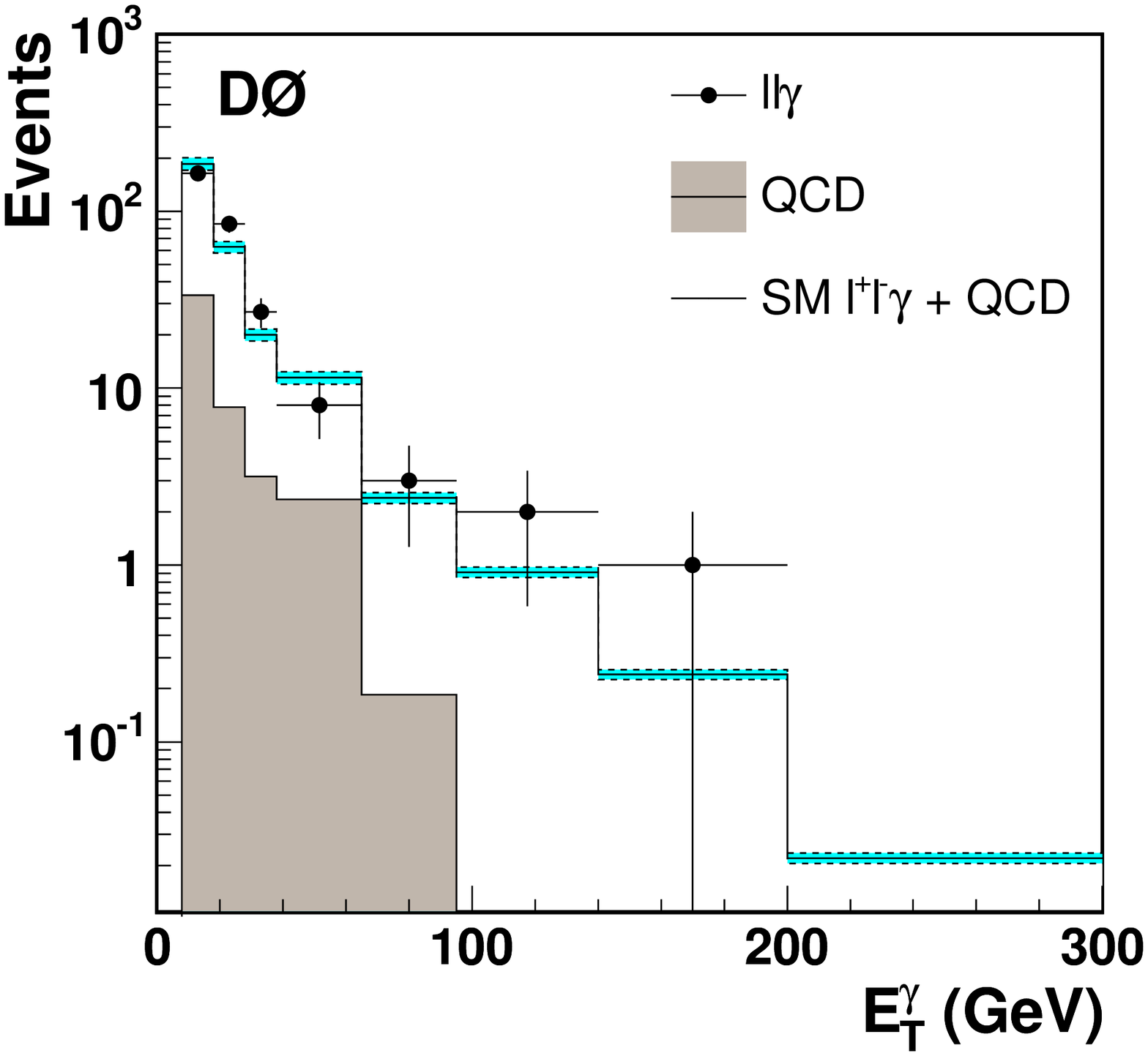,height=6.0cm}}
\end{picture}
\caption{The photon $E_T$ spectrum of the candidate events for the $Z \gamma$ candidates for CDF (left) and D$\emptyset$ (right). The data (black points) are 
compared with the SM expectations for signal and backgrounds (open histogram).
\label{figure:zg_spectra}}
\end{center}
\end{figure}

\begin{table}[htb]
\caption{One-dimensional 95\% C.L. limits on the anomalous $Z\gamma$ couplings
for $\Lambda~=~750$~GeV and 1.0~TeV.} 
\label{table:zzg_limits}
\vspace{0.4cm}
\begin{center}
\begin{tabular}{|c|c|c|}
\hline
Coupling & $\Lambda~=~750$~GeV & $\Lambda~=~1.0$~TeV   \\ \hline
$|\Re e(h^Z_{10,30})|, |\Im m(h^Z_{10,30})|$ & 0.24 & 0.23 \\
$|\Re e(h^Z_{20,40})|, |\Im m(h^Z_{20,40})|$ & 0.027 & 0.020 \\
$|\Re e(h^{\gamma}_{10,30})|, |\Im m(h^{\gamma}_{10,30})|$ & 0.29 & 0.23 \\
$|\Re e(h^{\gamma}_{20,40})|, |\Im m(h^{\gamma}_{20,40})|$ & 0.030 & 0.019  \\ \hline
\end{tabular}
\end{center}
\end{table}

Fig.~\ref{figure:zg_spectra} shows the photon $E_T$ spectrum for CDF (left) and D$\emptyset$ (right). The data (black points) are compared with the SM expectations for signal and backgrounds (open histogram). The data are in good agreement with the SM expectations, and no enhancement of the photon $E_T$ spectrum is seen at high transverse energy.

\begin{figure}
\setlength{\unitlength}{1.0cm}
\vskip 2.5cm
\begin{center}
\begin{picture}(3.5,4.5)
\put(-5.0,0.0){\epsfig{figure=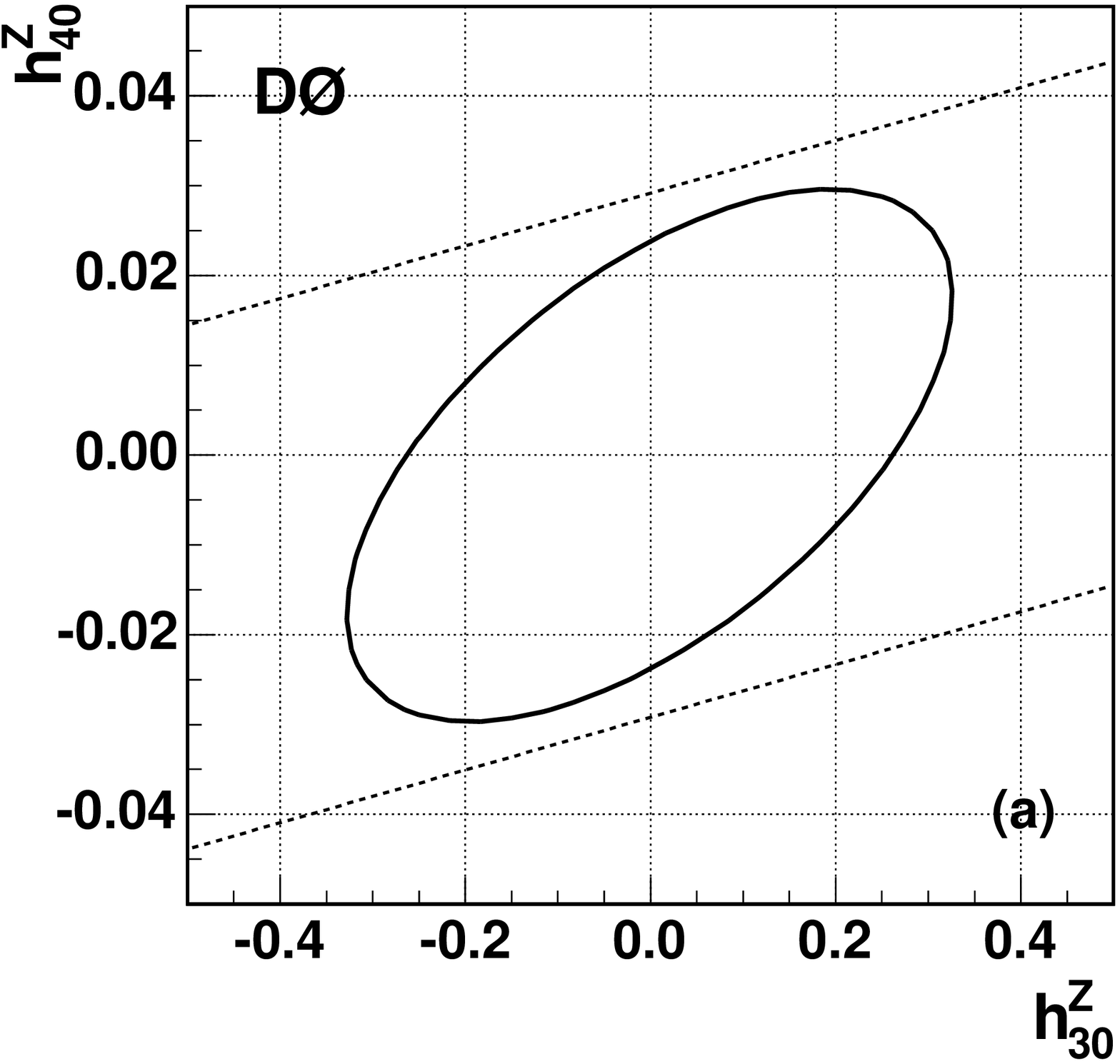,height=6.0cm}}

\put(2.5,0.0){\epsfig{figure=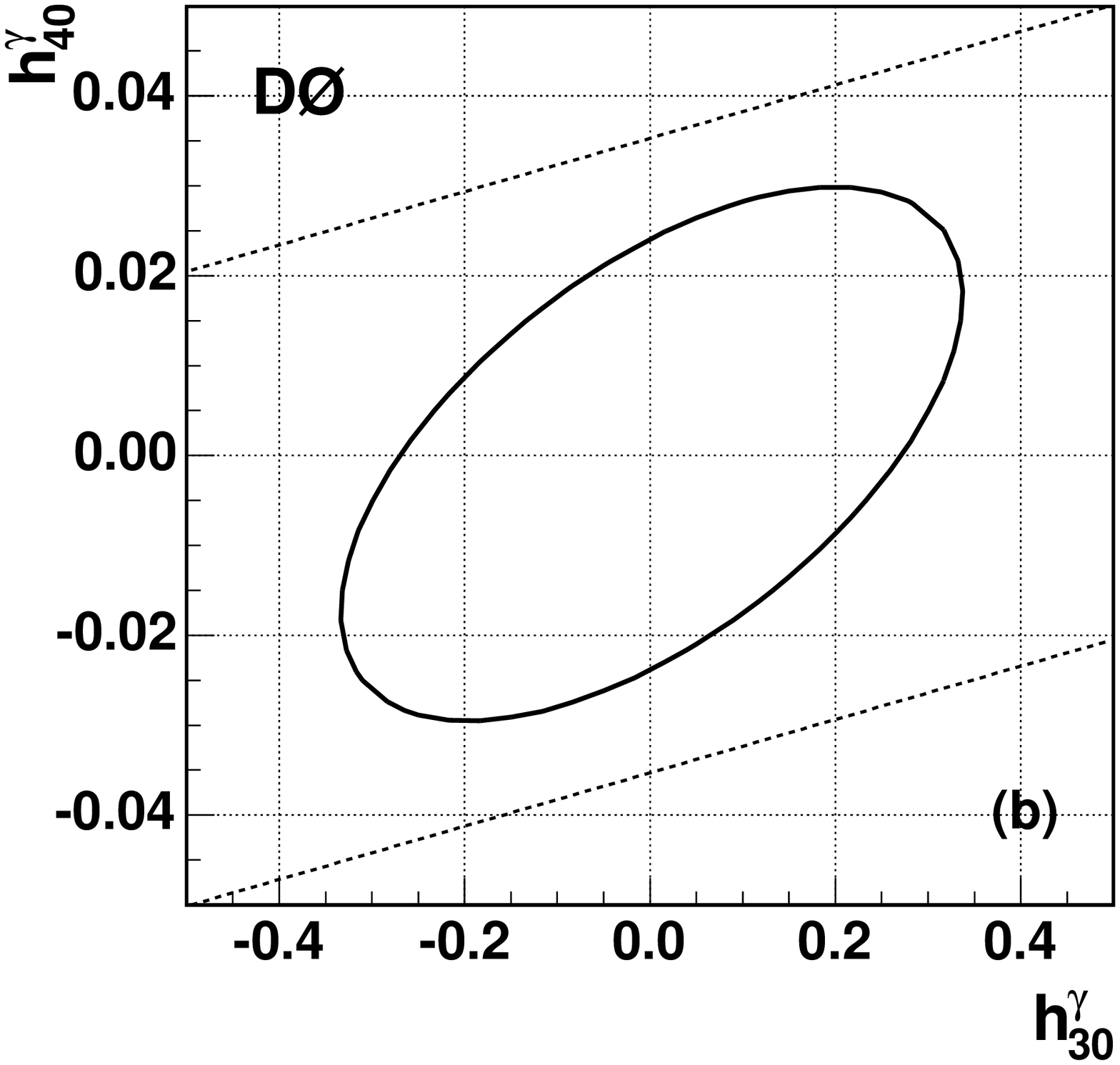,height=6.0cm}}
\end{picture}
\end{center}
\caption{Two-dimensional 95\% C.L. limits on CP-conserving $ZZ\gamma$ (a) and $Z\gamma\gamma$ (b) couplings for $\Lambda$~=~1~TeV obtained with 
286-324~pb$^{-1}$ of D$\emptyset$ data. Dashed lines illustrate the unitarity constraints.\label{figure:zzg_2d_limits}}
\end{figure}

From the $E^{\gamma}_T$ spectrum, 95\% C.L. limits are extracted by D$\emptyset$ on each of the CP-violating and CP-conserving anomalous couplings. The one-dimensional limits at 95\% C.L. on the real and imaginary parts of $h^V_{i0}$ ($i~=~1,2,3,4$) are given in Table~\ref{table:zzg_limits}.
The two-dimensional limit contours in the plane $h^V_{30}$ and $h^V_{40}$ at 95\% 
C.L. are shown in 
Fig.~\ref{figure:zzg_2d_limits}. These limits are substantially more restrictive than previous results which have been presented using the same formalism~\cite{D098}. The limits on $h^V_{20}$ and $h^V_{40}$ are more than twice as restrictive as the combined results of the four LEP experiments~\cite{LEP04}.

\section{Conclusion}
The Run~II of the Tevatron is well underway and the di-boson cross section 
measurements have already re-established most of the Run~I results. 
In addition limits on the $WZ$ cross section are extracted by both experiments
as well as limits on the triple gauge boson couplings.

\end{document}